\documentclass[aps,prd,twocolumn,superscriptaddress,showpacs,amsmath,amssymb]{revtex4-1}

\usepackage{graphicx}

\begin{document}


\title{Search for Dark Matter from the Galactic Halo with the IceCube Neutrino Observatory}



\affiliation{III. Physikalisches Institut, RWTH Aachen University, D-52056 Aachen, Germany}
\affiliation{Dept.~of Physics and Astronomy, University of Alabama, Tuscaloosa, AL 35487, USA}
\affiliation{Dept.~of Physics and Astronomy, University of Alaska Anchorage, 3211 Providence Dr., Anchorage, AK 99508, USA}
\affiliation{CTSPS, Clark-Atlanta University, Atlanta, GA 30314, USA}
\affiliation{School of Physics and Center for Relativistic Astrophysics, Georgia Institute of Technology, Atlanta, GA 30332, USA}
\affiliation{Dept.~of Physics, Southern University, Baton Rouge, LA 70813, USA}
\affiliation{Dept.~of Physics, University of California, Berkeley, CA 94720, USA}
\affiliation{Lawrence Berkeley National Laboratory, Berkeley, CA 94720, USA}
\affiliation{Institut f\"ur Physik, Humboldt-Universit\"at zu Berlin, D-12489 Berlin, Germany}
\affiliation{Fakult\"at f\"ur Physik \& Astronomie, Ruhr-Universit\"at Bochum, D-44780 Bochum, Germany}
\affiliation{Physikalisches Institut, Universit\"at Bonn, Nussallee 12, D-53115 Bonn, Germany}
\affiliation{Dept.~of Physics, University of the West Indies, Cave Hill Campus, Bridgetown BB11000, Barbados}
\affiliation{Universit\'e Libre de Bruxelles, Science Faculty CP230, B-1050 Brussels, Belgium}
\affiliation{Vrije Universiteit Brussel, Dienst ELEM, B-1050 Brussels, Belgium}
\affiliation{Dept.~of Physics, Chiba University, Chiba 263-8522, Japan}
\affiliation{Dept.~of Physics and Astronomy, University of Canterbury, Private Bag 4800, Christchurch, New Zealand}
\affiliation{Dept.~of Physics, University of Maryland, College Park, MD 20742, USA}
\affiliation{Dept.~of Physics and Center for Cosmology and Astro-Particle Physics, Ohio State University, Columbus, OH 43210, USA}
\affiliation{Dept.~of Astronomy, Ohio State University, Columbus, OH 43210, USA}
\affiliation{Dept.~of Physics, TU Dortmund University, D-44221 Dortmund, Germany}
\affiliation{Dept.~of Physics, University of Alberta, Edmonton, Alberta, Canada T6G 2G7}
\affiliation{Dept.~of Subatomic and Radiation Physics, University of Gent, B-9000 Gent, Belgium}
\affiliation{Max-Planck-Institut f\"ur Kernphysik, D-69177 Heidelberg, Germany}
\affiliation{Dept.~of Physics and Astronomy, University of California, Irvine, CA 92697, USA}
\affiliation{Laboratory for High Energy Physics, \'Ecole Polytechnique F\'ed\'erale, CH-1015 Lausanne, Switzerland}
\affiliation{Dept.~of Physics and Astronomy, University of Kansas, Lawrence, KS 66045, USA}
\affiliation{Dept.~of Astronomy, University of Wisconsin, Madison, WI 53706, USA}
\affiliation{Dept.~of Physics, University of Wisconsin, Madison, WI 53706, USA}
\affiliation{Institute of Physics, University of Mainz, Staudinger Weg 7, D-55099 Mainz, Germany}
\affiliation{Universit\'e de Mons, 7000 Mons, Belgium}
\affiliation{Bartol Research Institute and Department of Physics and Astronomy, University of Delaware, Newark, DE 19716, USA}
\affiliation{Dept.~of Physics, University of Oxford, 1 Keble Road, Oxford OX1 3NP, UK}
\affiliation{Dept.~of Physics, University of Wisconsin, River Falls, WI 54022, USA}
\affiliation{Oskar Klein Centre and Dept.~of Physics, Stockholm University, SE-10691 Stockholm, Sweden}
\affiliation{Dept.~of Astronomy and Astrophysics, Pennsylvania State University, University Park, PA 16802, USA}
\affiliation{Dept.~of Physics, Pennsylvania State University, University Park, PA 16802, USA}
\affiliation{Dept.~of Physics and Astronomy, Uppsala University, Box 516, S-75120 Uppsala, Sweden}
\affiliation{Dept.~of Physics, University of Wuppertal, D-42119 Wuppertal, Germany}
\affiliation{DESY, D-15735 Zeuthen, Germany}

\author{R.~Abbasi}
\affiliation{Dept.~of Physics, University of Wisconsin, Madison, WI 53706, USA}
\author{Y.~Abdou}
\affiliation{Dept.~of Subatomic and Radiation Physics, University of Gent, B-9000 Gent, Belgium}
\author{T.~Abu-Zayyad}
\affiliation{Dept.~of Physics, University of Wisconsin, River Falls, WI 54022, USA}
\author{J.~Adams}
\affiliation{Dept.~of Physics and Astronomy, University of Canterbury, Private Bag 4800, Christchurch, New Zealand}
\author{J.~A.~Aguilar}
\affiliation{Dept.~of Physics, University of Wisconsin, Madison, WI 53706, USA}
\author{M.~Ahlers}
\affiliation{Dept.~of Physics, University of Oxford, 1 Keble Road, Oxford OX1 3NP, UK}
\author{K.~Andeen}
\affiliation{Dept.~of Physics, University of Wisconsin, Madison, WI 53706, USA}
\author{J.~Auffenberg}
\affiliation{Dept.~of Physics, University of Wuppertal, D-42119 Wuppertal, Germany}
\author{X.~Bai}
\affiliation{Bartol Research Institute and Department of Physics and Astronomy, University of Delaware, Newark, DE 19716, USA}
\author{M.~Baker}
\affiliation{Dept.~of Physics, University of Wisconsin, Madison, WI 53706, USA}
% ...

\date{\today}

\collaboration{IceCube Collaboration}
\noaffiliation



\begin{abstract}
Self-annihilating or decaying dark matter in the Galactic halo might produce
high energy neutrinos detectable with neutrino telescopes. We have conducted a
search for such a signal using 276~days of data from the 
IceCube 22-string configuration detector acquired during 2007 and 2008.
The effect of halo model choice in the extracted limit is reduced by performing a search that considers the outer halo region and not the Galactic Center.
We constrain any large scale neutrino anisotropy and are able to set a limit on the dark matter self-annihilation cross section of $\langle \sigma_{A} v \rangle \simeq 10^{-22}{\rm cm}^3 {\rm s}^{-1}$ for WIMP masses above 1~TeV, assuming a monochromatic neutrino line spectrum.
\end{abstract}

\pacs{95.35.+d,98.35.Gi,95.85.Ry}
%
%

\maketitle

\section{Introduction}

There is compelling observational evidence for the existence of dark matter. 
Although knowledge of its underlying nature remains elusive, a variety of theories provide candidate particles~\cite{Bertone:2004pz}. 
Among those are Supersymmetry~\cite{Martin:1997ns} and 
Universal Extra Dimensions~\cite{Appelquist:2000nn}, both of which predict new physics at the electro-weak scale and, in most scenarios, introduce a 
light, and stable (or long lived) particle that exhibits the properties of 
a Weakly Interacting Massive Particle (WIMP)~\cite{Steigman:1984ac}. 
WIMPs are an ideal dark matter candidate, predicted to have masses ranging from a few tens of GeV to several TeV. 
High energy neutrinos are expected to be produced as a result of the self-annihilation or decay of WIMPs.
These neutrinos are detectable by high energy neutrino telescopes, making them 
powerful tools in the search for WIMPs and the investigation of their properties. 
In particular, they can be used to probe the self-annihilation cross section of dark matter candidates
by looking for anomalous neutrino signals from the Galactic halo.
Additionally, WIMPs could also be gravitationally captured by massive bodies like the Sun. If the annihilation rate of these captured WIMPs is regulated by the capture rate, then neutrino telescopes can be used to probe the WIMP-nucleon scattering cross section~\cite{Abbasi:2009uz}. 

Recent observations of a GeV positron
excess by PAMELA~\cite{Adriani:2008zr}, an anomalous electron peak by ATIC~\cite{ATIC:2008zzr}, and
electron spectra from H.E.S.S.~\cite{Aharonian:2009ah} and Fermi~\cite{Abdo:2009zk}, demonstrate the importance of a multi-messenger approach 
to astrophysics and validate the interest in a neutrino channel. 
The observed lepton signals are inconsistent with each other or standard electron--positron production models~\cite{Moskalenko:1997gh} and although they could potentially originate from nearby astrophysical sources (e.g. pulsars~\cite{Yuksel:2008rf}), they could also be an indication of dark matter.
If interpreted as the latter, it would suggest the existence 
of a leptophilic dark matter particle in the TeV mass range~\cite{Meade:2009iu,Cirelli:2008pk}. Such a model would also result in significant high energy neutrino fluxes, through the decay of muons and $\tau$-leptons.
A significant fraction of neutrinos could also be produced directly as part of the annihilation~\cite{Lindner:2010rr}, producing a line feature in the resulting neutrino spectrum. 
Such a mono-energetic neutrino flux is of specific interest since it can be used to set a model 
independent limit on the total dark matter self-annihilation cross section~\cite{Beacom:2006tt} for the
region of parameter space where gamma-ray signals would dominate.

In this paper we discuss a search for neutrino signals produced by annihilating or decaying dark matter in the Galactic halo. 
The search is used to test the self-annihilation cross section by constraining the product of cross section and velocity averaged over the dark matter velocity distribution, $\langle \sigma_{A} v \rangle$, and to probe the lifetime, $\tau$.

The search focuses on the outer Milky Way halo, where the dark matter density distributions are relatively well modelled. We do not include the Galactic Center region and thus remove any strong dependence on the choice of the halo profile. 
We quantify the residual weak dependence and present constraints on the dark matter self-annihilation 
cross section and lifetime in a model-independent way for a set of selected benchmark annihilation and decay channels, respectively.

The paper is organized as follows: in the next section we describe the detector used for the data taken during 2007--2008 which is the base for our analysis. 
Section III discusses how we obtain an expected neutrino flux at Earth using different dark matter distributions and annihilation channels. In section IV we describe our data selection criteria and analysis strategy, which is followed by a discussion of the associated systematic uncertainties in section V. Section VI presents the result of the search, and section VII puts it in context with other experiments. Section VIII concludes by summarizing the results and giving an outlook for related searches.

\section{The IceCube Neutrino Observatory and Event Selection}

\begin{figure}[htb]
\includegraphics[width=3.0in]{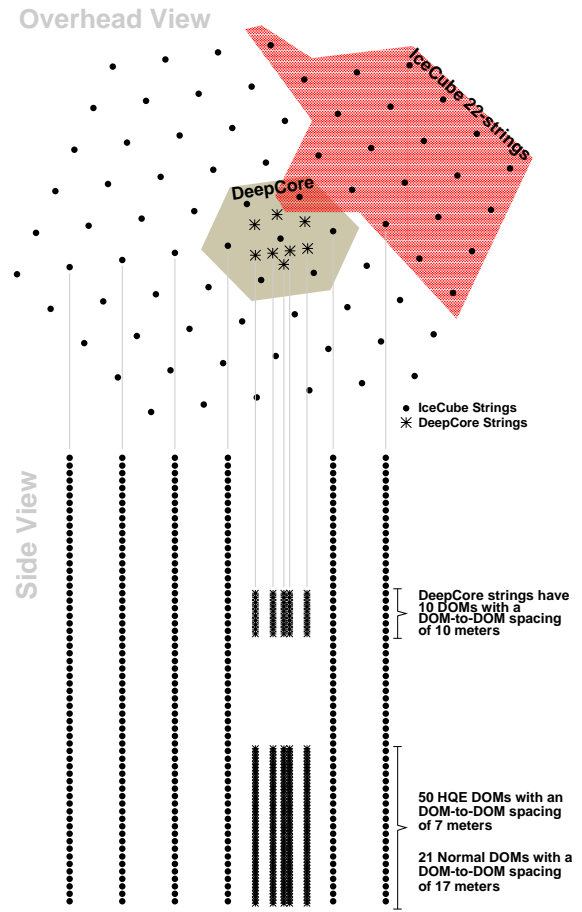}
\caption{(Color online) Schematic view of the IceCube Neutrino Observatory including the 
low energy extension DeepCore.
Shown in red is the partially instrumented detector, active in the 2007--2008 season, which was the only portion used for this analysis.
\label{fig_icecube_schematic}}
\end{figure}

The IceCube Neutrino Observatory, located at the geographic South Pole, consists of 
the IceCube neutrino telescope and the IceTop air shower array~\cite{Achterberg:2006md}. 
IceTop covers a surface area of one square-kilometer above the IceCube ``in-ice'' detector, and is designed to measure cosmic ray air 
showers with energies between approximately 300~TeV to 1~EeV. 
The in-ice detector instruments a volume of one
cubic kilometer of Antarctic ice~\cite{IceCube:icepaper} with 5160~digital optical modules (DOMs)~\cite{Abbasi:2008ym}
deployed at depths between 1450~m and 2450~m (see Fig.~\ref{fig_icecube_schematic}).
The DOMS are distributed over 86 electrical cable bundles (strings) that handle power 
transmission and communication with electronics located on the surface.
Each DOM consists of a 25~cm Hamamatsu R7081-02 photomultiplier tube~\cite{Abbasi:2010vc} connected to a waveform recording data 
acquisition circuit.
It is capable of resolving pulses with nanosecond precision and has an effective dynamic range of 200~photoelectrons per 15~ns.

IceCube detects all flavors of active neutrinos through Cherenkov light emission from secondary particles created when a neutrino interacts in the ice. 
Muon neutrinos are of particular interest 
since their extended track-like topology makes them relatively simple to identify.
Furthermore, the elongated tracks of the muons 
permit a relatively accurate reconstruction of the neutrino direction with approximately a few degrees precision at the detection threshold of 100~GeV.
Neutrinos with energies down to about 10~GeV can be identified in a densely
instrumented sub-detector, DeepCore~\cite{Wiebusch:2009jf}, which has been operating since early 2010 (see Fig.~\ref{fig_icecube_schematic}). In this analysis, we use data taken with an intermediate construction stage of the in-ice detector, comprising 22 strings.

The primary background in the search for neutrinos originates from cosmic ray air showers.
When high energy cosmic rays enter the Earth's upper atmosphere they produce extended air showers, a part of which includes high energy pions and kaons.
The decay of these mesons results in a continuous stream of
neutrinos and muons. These are known as atmospheric muons and neutrinos, and their
flux is regulated by the path length and time the parent particles had in the 
atmosphere to lose energy or decay.
The resulting neutrino spectrum obeys a power law with a spectral index of $\gamma \approx 3.7$~\cite{Honda:1995hz,Agrawal:1995gk}. 
High energy muons are capable of travelling long distances through matter before they eventually decay, resulting in a down-going muon flux at the IceCube detector.
In contrast, neutrinos below 100~TeV can traverse the Earth without significant absorption losses.
To distinguish between a muon produced from a charged current interaction of a muon neutrino from those produced in the atmosphere, we select only tracks 
that enter the detector from below the horizon.
Given the 22-string detector configuration (see Fig.~\ref{fig_icecube_schematic}) for the analysis presented here, 
the total trigger rate was approximately $550$~Hz, dominated by down-going atmospheric muons.  A pre-selection at the South Pole for up-going reconstructed tracks reduces the data rate to 20~Hz, which is sent by satellite to be processed offline.

\section{Halo Profiles and Signal Expectations}

\begin{figure}[htb]
\includegraphics[angle=-90,width=3.0in]{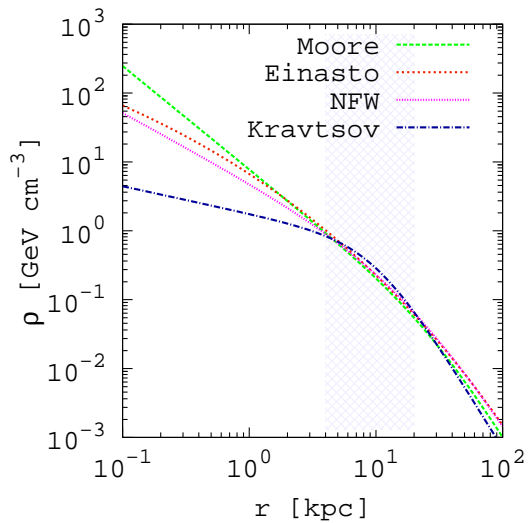}
\caption{Comparison of the dark matter density distribution, $\rho(r)$, as a
  function of distance from the Galactic Center as
  described by the Einasto, NFW, Kravtsov, and Moore halo profiles. The shaded area indicates the region where the presented analysis is sensitive.  
\label{fig_halo_profiles}}
\end{figure}

Recent advances in N-body simulations~\cite{Diemand:2007qr} and gravitational lensing observations~\cite{Menard:2009yb} have provided reliable predictions of the dark matter density distribution in
the Milky Way~(MW). While the outer regions of the dark matter
halo of the Milky Way (several kpc away from the Galactic Center (GC)) are
relatively well modelled, the structure of its central region is
still a matter of debate since it can neither be resolved in simulations,
nor directly measured. Not surprisingly, halo models generally show very
similar behavior at large distances from the Galactic Center, but differ
significantly in their predictions near it.
This effect is shown in Fig.~\ref{fig_halo_profiles}, where the dark matter density, $\rho(r)$, predictions from several spherically symmetric halo profiles obtained from N-body simulations are compared. 
We show four different distribution functions which are used in our analysis.  
Since we only use neutrinos from the northern sky, the effective dark matter densities which dominate the analysis are those between a distance from roughly 4~kpc to 20~kpc from the Galactic Center. In this range the various halo profiles are relatively consistent in their description of the dark matter density. This agreement allows us to constrain the dark matter 
self-annihilation cross section with minimal halo profile dependence.

We use the Einasto~\cite{Einasto_profile,Einasto:by_hand} and 
Navarro-Frenk-White (NFW)~\cite{Navarro:1995iw} profiles as benchmark models, while the Moore~\cite{Moore:1999gc}, and
Kravtsov~\cite{Kravtsov:1997dp} profiles are applied as extreme cases to estimate the
impact of halo model choice on the result.
The Einasto profile is given by:

\begin{equation}
\rho(r) = \rho_{-2} \times e^{(-\frac{2}{\alpha})\left[\left(\frac{r}{r_{-2}}\right)^{\alpha}-1\right]}
\end{equation}
with $\alpha=0.16$~\cite{Navarro:2003ew}, $r_{-2}=20$~kpc, and $\rho_{-2}$ normalized to a dark matter density $0.3 \frac{\rm GeV}{\rm
  cm^3}$ at the solar system's orbit in the Galaxy ($R_{\rm sc}$ = 8.5~kpc).
The remaining three profiles can
be described by the following function: 

\begin{equation}
\rho(r) = \frac{\rho_{0}}{(\frac{r}{r_s})^{\gamma}\left[1+(\frac{r}{r_s})^{\alpha}\right]^{(\beta-\gamma)/\alpha}},
\label{eq:dm_dist}
\end{equation}
where the variables $(\alpha, \beta, \gamma, r_{s})$~\cite{Yuksel:2007ac} take different numerical values (listed in Table~\ref{table:Halo_Parameters}) for the three models. The normalizations
are chosen such that the mass contained within the orbit of the Solar System in the Galaxy 
provides the appropriate dark matter contribution to the local rotational curves, and yields a local
dark matter density $\rho_{\rm sc}=\rho_{\rm NFW}(R_{\rm sc}) = 0.3 \frac{\rm GeV}{\rm
  cm^3}$ for the NFW profile.

\begin{table} 
\caption{Summary of the parameters of Eq.~\ref{eq:dm_dist} used in this analysis.\label{table:Halo_Parameters}}
\begin{ruledtabular}
\begin{tabular}{|l|r|r|r|r|r|}
Halo Model & $\alpha$ & $\beta$ & $\gamma$ & $r_{s}$/kpc & $\rho(R_{\rm sc})$/$\frac{\rm GeV}{\rm cm^3}$  \\
\hline
NFW	 & 1	& 3	& 1 	& 20	& 0.3 \\
Moore	 & 1.5	& 3	& 1.5 	& 28	& 0.27 \\
Kravtsov & 2	& 3	& 0.4 	& 10	& 0.37 \\
\end{tabular}
\end{ruledtabular}
\end{table}

The expected neutrino flux, $\phi_{\nu}$, from dark matter self-annihilations is proportional
to the square of the dark matter density integrated along the line of sight
$J(\psi)$:

\begin{equation}
J(\psi) = \int_{0}^{l_{\rm max}}
\frac{\rho^2(\sqrt{R_{\rm sc}^2 - 2 l R_{\rm sc} \cos\psi + l^2})}{R_{\rm sc} \rho^2_{\rm sc}}  dl,
\label{Jpsi_integral}
\end{equation}
where $\psi$ is the angular distance from the Galactic Center and $l_{\rm max}$ is the upper limit of the integral, defined as

\begin{equation}
l_{\rm max} = \sqrt{(R^2_{\rm MW}-\sin^2\psi R^2_{\rm sc})}+R_{\rm sc}\cos\psi.
\end{equation} 

We adopt a halo size of $R_{\rm MW}=40$~kpc. Contributions to the expected neutrino flux from beyond this range are
small, and are discussed as part of our systematic studies on the result in section VI. 

\begin{figure}
\includegraphics[angle=-90,width=2.8in]{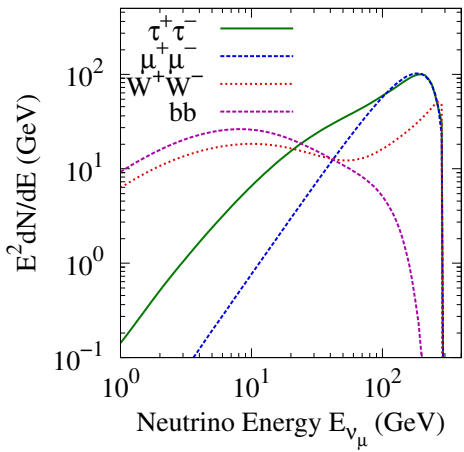}
\caption{Differential muon neutrino energy spectrum per annihilation, taking neutrino oscillations into account.
In this example we assume a WIMP mass of $300$~GeV and 100\% branching fraction into the corresponding annihilation channel.
\label{fig_Halo_Multi_dNdE_log}}
\end{figure}

The annihilation products are highly model dependent
and we thus study extremes of the possible annihilation channels assuming a branching ratio of 100\% for each of them in turn.
We consider soft neutrino spectra produced from the annihilation into quarks 
(${\rm b}\bar{\rm b}$), and hard spectra as produced by annihilation into 
${\rm W}^+{\rm W}^-$ and $\mu^{+}\mu^{-}$.
In addition, we consider a neutrino line spectrum ($\chi\chi \rightarrow \nu \nu$).

Neutrinos will have undergone extensive mixing through vacuum oscillations over the distances travelled across the Galaxy.
We determine 
neutrino flavor oscillations in the long baseline limit~\cite{PhysRevD.67.073024,Murase:2007yt},
adopting values of $\sin^{2}2\Theta_{12} = 0.86$, $\Theta_{23}$ maximal $(\Theta_{23} \simeq \pi/4)$, and $\Theta_{13} \simeq 0$.
The neutrino fluxes at Earth are then given by:

\begin{equation}
\phi_{\nu_{e}} \simeq \phi_{\nu_{e}}^{0} - \frac{1}{4} s_2
\end{equation}
and 

\begin{equation}
\phi_{\nu_{\tau}} \simeq \phi_{\nu_{\mu}} \simeq \frac{1}{2}(\phi_{\nu_{\mu}}^{0} + \phi_{\nu_\tau}^{0}) + \frac{1}{8} s_2,
\end{equation}
where $\phi_{\nu_{i}}^{0}$ is the flux at injection and $s_2$ is defined as 
$ \sin^2 2\Theta_{12} (2\phi_{\nu_{e}}^{0} - \phi_{\nu_{\mu}}^{0} - \phi_{\nu_{\tau}}^{0} )$. 
Note that the expected flux for muon and tau neutrinos is equal.

The neutrino energy spectra
were produced using DarkSUSY~\cite{Gondolo:2004sc}, an advanced numerical software package for supersymmetric dark matter calculations, 
and are shown in Fig.~\ref{fig_Halo_Multi_dNdE_log}. 

The differential neutrino flux from the annihilations of neutralinos of mass
$m_{\chi}$ in the Galactic halo is given by~\cite{Yuksel:2007ac}:

\begin{equation}
\frac{d\phi_{\nu}}{dE} = \frac{\langle \sigma_A v \rangle}{2} J(\psi) \frac{R_{\rm sc}
  \rho_{\rm sc}^2}{4\pi m^2_{\chi}} \frac{dN_{\nu}}{dE},
\end{equation}
where $\frac{dN_{\nu}}{dE}$ is the differential neutrino multiplicity per annihilation.
Similar to the annihilation cross section, one can search for signals from
decaying dark matter~\cite{PalomaresRuiz:2010pn} and constrain 
the lifetime, $\tau$.  
For decaying dark matter, the expected neutrino flux is proportional to the dark matter density along the
line of sight, given by:

\begin{equation}
J_{\rm d}(\psi) = \int_{0}^{l_{\rm max}}
\frac{\rho (\sqrt{R_{\rm sc}^2 - 2 l R_{\rm sc} \cos\psi + l^2})}{R_{\rm sc} \rho_{\rm sc}} dl.
\label{eq_line_of_sight}
\end{equation}

The expected neutrino flux from the dark matter decay is then:

\begin{equation}
\frac{d\phi_{\nu}}{dE} = \frac{1}{\tau} J_{\rm d}(\psi) \frac{R_{\rm sc}\rho_{\rm
    sc}}{4\pi m_{\chi}} \frac{dN_{\nu}}{dE}.
\label{dphidE_decay}
\end{equation}

We use identical halo model parameters in both the dark matter annihilation and decay analyses.
We assume a smooth halo
profile and discuss the effect of substructure separately.

\section{Data Selection}

The search for a clustering of neutrinos to indicate an astrophysical neutrino source is one of the benchmark analyses performed by the IceCube collaboration. 
Such a ``point source'' search relies on muon neutrinos since the elongated tracks of the muons permit an accurate reconstruction of the neutrino direction. The 22-string detector configuration has 
produced a well understood neutrino candidate sample~\cite{Abbasi:2009iv}, extracted 
using likelihood-based track
reconstructions and selecting tracks from $-5^\circ$ to $85^\circ$ in declination.
The shape of the likelihood function around the best-fit value is used to estimate the angular uncertainty of the reconstructed track~\cite{Neunhoffer:2004ha}, while the number of optical modules in the event which record minimally scattered Cherenkov photons gives an additional handle on the quality of the reconstruction.
Such ``direct'' photons are isolated via a time difference selection window between the expected arrival time of an unscattered
photon, given the reconstructed track, and the registered DOM hit time.  
Near the horizon, the background from poorly reconstructed atmospheric muons is further reduced by 
an additional cut on the likelihood ratio of the best-fit track to the best-fit track constrained to be down-going.
These applied selection criteria remove the largest fraction of mis-reconstructed
down-going events, maintaining a neutrino candidate sample with about $90\%$ purity~\cite{Abbasi:2009iv}.
The final northern sky dataset consists of $5114$~neutrino candidate events acquired in $275.7$~days of livetime. 

\begin{figure}[htb]
\includegraphics[width=3.0in]{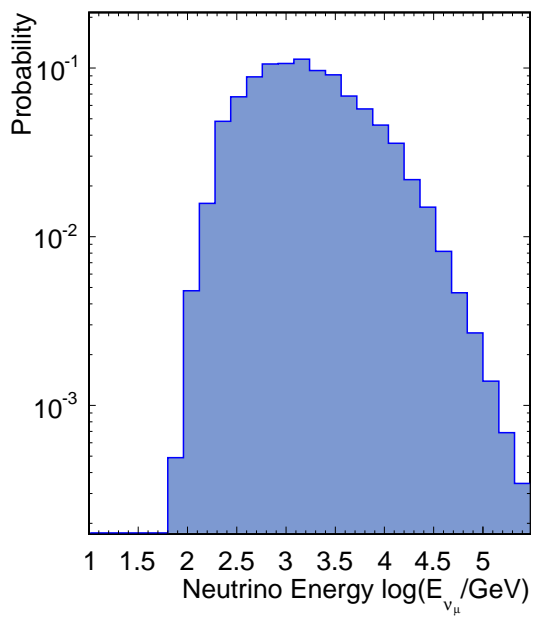}
\caption{Muon neutrino energy distribution from atmospheric neutrino simulations at final selection level. \label{fig_nu_energy}}
\end{figure}

Figure~\ref{fig_nu_energy} shows the neutrino energy distribution of
the final selection based on simulations of atmospheric neutrinos.

\begin{figure}
\includegraphics[width=3.0in,height=2.5in]{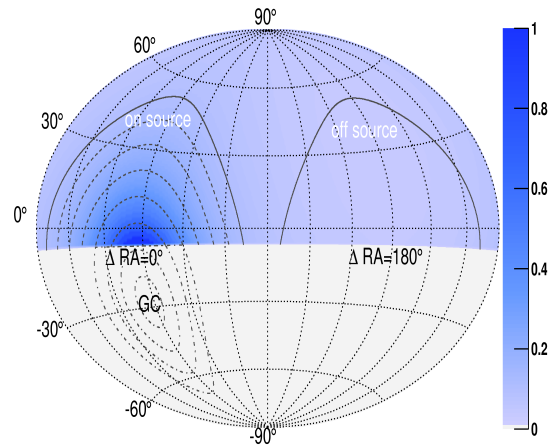}
\caption{The relative expected neutrino flux from dark matter self-annihilation in the northern celestial hemisphere of the
  Milky Way Galaxy halo is shown. The largest flux is 
  expected at a right ascension (RA) closest to the
  Galactic Center ($\Delta {\rm RA} =0$). 
  Dashed lines indicate circles around the Galactic Center with a half-opening angle, $\psi$, that increases in $10^\circ$ steps. The solid
  lines show the definition of on-- and off--source regions in the northern
  hemisphere. The on--source region is centered around $\Delta {\rm RA} =0$, while
  the off--source region is shifted by $180^{\circ}$ in RA. 
\label{fig_OnOffSourceRegion}
}
\end{figure}

Assuming a given annihilation channel and dark matter halo profile, one can determine the expected neutrino flux (proportional to the dark matter annihilation cross section) for any given location on the sky.
The flux is peaked in the direction of the Galactic Center, which is a prominent 
target for searches. However, the Galactic Center is located in the southern hemisphere at $266^\circ$~right ascension (RA) and
$-29^\circ$~declination (DEC), and therefore outside the field of view in the used dataset.

In the northern hemisphere, regardless of the choice of halo model, dark matter annihilations 
would produce a large--scale neutrino anisotropy.
The search for such an anisotropy affords distinct advantages 
for discovery. An observation of a flux from the Galactic Center 
would be more difficult to
distinguish from other astrophysical sources 
or cosmic ray interaction with the interstellar medium. 
However, the Galactic Center is an excellent target to constrain the dark 
matter self-annihilation cross section for a given halo model and 
is the subject of a separate analysis.

To test for an excess flux of neutrinos, we define two regions on the northern sky.
The first region will serve as our signal region (on--source) and is defined by a half-opening angle, $r_{\psi}$, around the Galactic Center. 
An equally sized region, offset by $180^\circ$ in RA, serves an off--source region (see Fig.~\ref{fig_OnOffSourceRegion}). 
This choice is motivated by the robustness and simplicity of the ensuing analysis and 
minimizes systematic uncertainties due to azimuth angle dependent reconstruction
efficiencies. For spherical halo profiles, the expected flux is a function of the angular distance from the Galactic Center, $\psi$, and we count the total number of
neutrino candidate events in each region. This makes the analysis maximally independent of
halo profiles and provides sensitivity to both hard and soft neutrino spectra.
 
The difference in the expected number of neutrino events between the on--source and off--source region is given by:

\begin{equation}
\Delta N = 
\left(N_{{\rm on}}^{\rm bkg}+N_{{\rm on}}^{\rm sig}\right) -
\left(N_{\rm off}^{\rm bkg}+N_{{\rm off}}^{\rm sig}\right)
\label{eq1},
\end{equation}
where bkg/sig stand for background and signal, respectively.
Background events are expected to be equally distributed in the on-- and off--source regions, simplifying the prediction to
$\Delta N^{\rm sig} = N_{{\rm on}}^{\rm sig} - N_{{\rm off}}^{\rm sig}$. 
The signal expectation in both regions, and hence $\Delta N^{\rm sig}$, is directly proportional to the dark matter self-annihilation cross-section $\langle \sigma_A v\rangle $.
To optimize the size of the on-- and off--source region, we 
chose an example cross section $\langle \sigma_{A} v\rangle _0$ and predict the expected 
number of signal events $S=\Delta N^{\rm sig}$ from simulations for 
different choices of $r_{\psi}$~\cite{Rott:2009hr}. 
For $r_{\psi} = 80^{\circ}$, the ratio of ${\rm S}/\sqrt{\rm B}$, where $B$ is the expected number of background events, is close to maximal for all considered halo profiles, while the on-- and off--source regions remain well separated and do not overlap.

\section{Systematic Uncertainties}

We first discuss the systematic uncertainty
associated with the background estimation.
By design, the
background can be determined from the data by comparing events in the on-- and off--source regions, eliminating most detector related effects.
Thus, only pre-existing anisotropy in the data must be considered.
The two dominant effects giving rise to this are: (1) An anisotropy in the cosmic ray
flux producing the atmospheric muon neutrino flux; 
(2) Variations in exposure for different RA. 

A large--scale anisotropy in the cosmic ray flux has been observed both on the
northern hemisphere by the TIBET air shower array~\cite{Amenomori:2006bx}, and
the southern hemisphere by an IceCube 
measurement of the down-going muon flux~\cite{Abbasi:2010mf}.
The northern hemisphere anisotropy for cosmic ray energies around 50~TeV is relevant to this analysis. This energy range of cosmic ray showers contributes most in creating this analysis' background up-going atmospheric muon neutrino flux. 
The overall scale of the measured cosmic ray anisotropy is about $0.2\%$, with peak values at ${\rm RA\approx}60^{\circ}$ and a minimum at  ${\rm RA\approx}180^{\circ}$. This is not aligned 
with an expected signal anisotropy from the Milky Way dark matter halo.
To provide a conservative systematic uncertainty estimate, we assume the
worst case of an aligned anisotropy, which peaks in one
region and is minimal in the other. In such a scenario
a difference of three events between on-- and off--source
regions would be observed, corresponding to a $0.2\%$ systematic uncertainty on the number of
background events.

\begin{figure}
\includegraphics[angle=-90,width=3.0in]{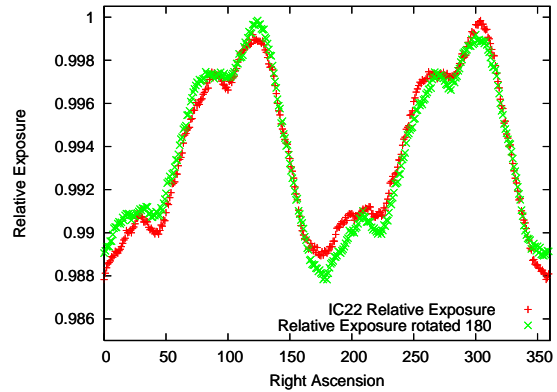}
\caption{
The relative exposure variation as function of RA and 
rotated by $180^{\circ}$ is shown. The absolute variation defines the signal 
acceptance uncertainty due to exposure, while the difference between the normal and rotated
exposure defines the corresponding systematic uncertainty on the background estimate.
\label{fig:exposure}}
\end{figure}

The muon track reconstruction efficiency varies as function of
the zenith angle and azimuth 
angle~\cite{Achterberg:2007bi,Abbasi:2009iv}. 
Although the azimuth dependence is relatively uniform for the axially symmetric full IceCube detector, it
is particularly pronounced in the partially instrumented 22-string detector configuration used for this analysis.
As the Earth rotates, each detector alignment in RA gets equal exposure within one sidereal day.
A small fraction of detector operations is dedicated towards scheduled detector maintenance, which is performed at times when communication with the South Pole can be established. The use of geosynchronous satellites introduces a bias in sidereal time, which means that fewer physics data runs are available for particular alignments of the detector in RA.
Selecting symmetric on-- and off--source
regions shifted by $180^{\circ}$ in RA 
reduces this effect significantly, such that  
the track reconstruction efficiency is almost identical to the case where the detector is rotated
by $180^{\circ}$.
The total expected variation in the number of events 
 due to this effect is approximately $0.1\%$ (see
Fig.~\ref{fig:exposure}).

It is possible, in principle, to correct for both the cosmic 
ray anisotropy and detector uptime effects. 
Because of their negligibly small impact compared to the background statistical uncertainty, such a correction has not been applied.
The contribution from the cosmic ray anisotropy ($0.2\%$) and the uneven
exposure ($0.1\%$) are uncorrelated.  
We use $0.3\%$ as a conservative estimate on the total systematic uncertainty 
on the number of background events in the on--source region (see Table~\ref{sys_summary}).

\begin{table}
\caption{Summary of systematic uncertainties on the background estimate.\label{sys_summary}}
\begin{ruledtabular}
\begin{tabular}{|l|r|}
Effect & Sys. Uncertainty \\
\hline
Cosmic ray anisotropy         & $0.2\%$ \\
Exposure                      & $0.1\%$ \\ 
\hline
Total Background              & $0.3\%$ \\
\end{tabular}
\end{ruledtabular}
\end{table}

The signal acceptance uncertainty is dominated by uncertainties in the ice properties and limitations in the detector simulation, which is uncorrelated with a number of theoretical uncertainties such as muon propagation, neutrino
cross section, and bedrock uncertainty, each of which have
been studied in previous analyses~\cite{Abbasi:2009iv}.
In addition, we consider the uncertainty due to Monte Carlo simulation statistics and detector exposure.
The individual track pointing uncertainty (point spread function), on the order of one degree, 
is negligible in this analysis, which targets a large--scale anisotropy.

Our dominant systematic uncertainty, the limited knowledge of ice properties
as a function of depth and limitations in the detector simulation, is expected to produce an 
observed discrepancy between data and simulation for events near the horizon~\cite{Abbasi:2009iv}.
For nearly horizontal tracks the disagreement is maximal, with $30\%$ more
events observed in data compared to simulation predictions. Since we use the data itself to predict the number of background events in the on--source region, this discrepancy does
not affect the background estimate.  
However, the signal acceptance can only be obtained from simulations. Hence, we must take this discrepancy into consideration for the signal acceptance uncertainty.
The higher than expected observed data rate, when compared to simulation expectations, may indicate 
a contribution from mis-reconstructed down-going events, or a
higher signal acceptance than expected. Both would cause
the constraints presented later to be more conservative.
The estimate for this systematic uncertainty in signal acceptance is 25-30\%.

The track reconstruction efficiency coupled with detector uptime (see
Fig.~\ref{fig:exposure}) results in a systematic
uncertainty on the signal acceptance of $1\%$. This uncertainty, combined with the 
theoretical uncertainties, results in a negligible 
contribution 
compared to the uncertainties in the optical properties of the ice.
We therefore assume a 30\% systematic signal acceptance uncertainty,
primarily associated with that from the ice properties and limitations in the detector simulation.

An additional systematic uncertainty to consider in signal acceptance is related to the photon detection efficiency of the DOMs, measured to be 8\% in the laboratory~\cite{Abbasi:2010vc}. 
The effect of this uncertainty on the passing rate of reconstructed tracks is found to range from about $1\%$ for energetic events ($\ge 1$~TeV), increasing to as much as 20\% for lower energy events ($\le 200$~GeV), as expected from annihilations assuming WIMP's of mass 200~GeV. We calculate this uncertainty for each of the considered WIMP masses and annihilation channels, then we add it in quadrature to the ice properties uncertainty discussed above.

To derive the total uncertainty on the signal acceptance, we have added the systematic signal acceptance uncertainty in quadrature to the statistical uncertainty (Monte Carlo statistics).
The Monte Carlo statistics uncertainty ranges from 3-6\% (hard channels) and 4-16\% (soft channels) in the TeV mass range dark matter, and increases to 50\% (hard channels) and 90\% (soft channel) at $m_{\chi}=200$~GeV.

\section{Results}

\begin{figure}
\includegraphics[width=3.0in]{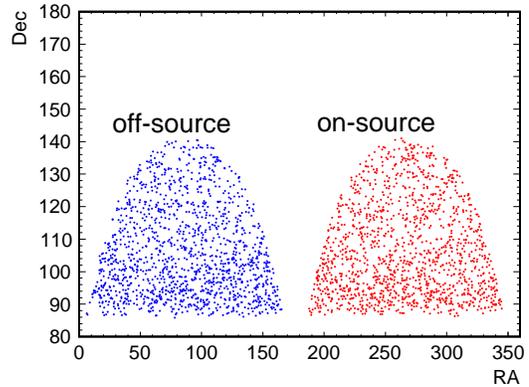}
\caption{
The location of the neutrino candidate
events in DEC versus RA for the on-- (right) and off--source (left) region.
\label{fig_on_off_source}}
\end{figure}

\begin{figure}
\includegraphics[width=3.0in]{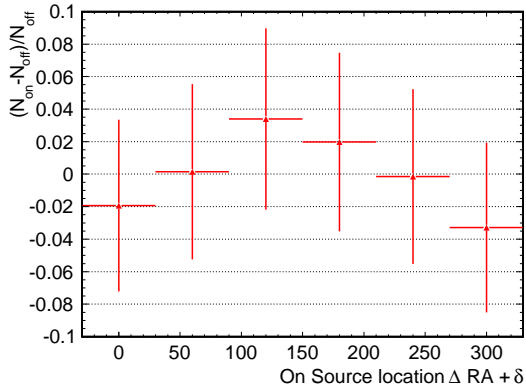}
\caption{Relative difference in number of events in the on/off--source region as a function of offset 
from the nominal position. The regions are shifted by $60^{\circ}$ steps to
be centered at $\Delta RA + \delta$. 
Error bars represent the statistical uncertainty in the bin.  
Adjacent bins are correlated, as regions partially overlap.
Note the 
first bin corresponds to the result obtained by this analysis. Bins 4-6 are closely related to bins 1-3, as $N_{\rm on}$ and $N_{\rm off}$ are swapped in them.
\label{fig_rotate60}}
\end{figure}

Except for examination of the data for quality assurance, 
the optimization of the size of the on--source region was performed 
entirely with simulated events, ensuring a blind analysis.
In the final dataset we observed 1389~events in the off--source 
region and 1367~events in the on--source region,
consistent with the null hypothesis. 
Figure~\ref{fig_on_off_source} shows the distribution of these neutrino candidate
events in declination and right ascension.
To study the possibility of an anisotropy in an adjacent bin, we shift the on-- and off--source regions in $60^{\circ}$ steps.
For each of the step bins, the ratio of ${\rm N}_{\rm on}/{\rm N}_{\rm off}$ is 
consistent with one (see Fig.~\ref{fig_rotate60}).

We compute constraints on the neutrino flux from
dark matter annihilation from the Galactic halo.
Given a specific $\langle \sigma_{A} v\rangle _{0}$ in signal simulations, the number of 
expected events for an arbitrary cross section $\langle \sigma_{A} v\rangle$ is

\begin{equation}
\Delta N^{\rm sig}(\langle \sigma_{A} v\rangle ) = \frac{\langle \sigma_{A}
  v\rangle }{\langle \sigma_{A} v\rangle _{0}} (\Delta N^{\rm sig}(\langle \sigma_{A} v\rangle _0)).
\end{equation}

The cross section limit at 90\%~C.L. is

\begin{equation}
\langle \sigma_{A} v\rangle _{90} = \Delta {\rm N}_{90} \times \frac{\langle \sigma_{A} v\rangle_{0}}{\Delta {\rm N}^{\rm sig}(\langle \sigma_{A} v\rangle_{0} )},
\end{equation} 
where $\Delta {\rm N}_{90}$ is the limit at 90\%~C.L. for the number of signal events.

To determine $\Delta {\rm N}_{90}$, we construct a Neyman confidence belt. 
The one-sided 90\%~C.L. acceptance intervals are determined by a simple Monte Carlo, in which 
the numbers of events in the on-- and off--source regions
are assumed to be Poisson distributed over repeated measurements, with an average contribution of $N_{\rm bkg} = N_{\rm off} = 1389 \pm 4 ({\rm sys}) 37 ({\rm stat})$. The 90\%~C.L. event upper limit $\Delta {\rm N}_{90}$ is calculated for various WIMP masses and annihilation channels using the appropriate signal expectation. Statistical and systematic uncertainties in the signal expectations are represented by log-normal distributions. For a 30\% signal acceptance uncertainty, for example, the upper limit was found to be $\Delta {\rm N}_{90}=49$ for $\Delta {\rm N} = -22$~events.
For small signal acceptance uncertainties, where the log-normal distribution can be approximated by a Gaussian, results are consistent with the confidence interval constructed using the method by Lundberg et al.~\cite{Lundberg2010683,Conrad:2002kn}.
Our limit calculation of the on--source region also resembles a
commonly used procedure by Li and Ma to compute the significance of an on--source observation~\cite{Li:1983fv}. 
The significance $\xi$ is defined as 

\begin{equation}
\xi = \frac{N_{\rm on}- \eta N_{\rm off}}{\eta\sqrt{N_{\rm on}+N_{\rm off}}} \approx \frac{\Delta N}{\sqrt{2\times N_{\rm off}}}.
\end{equation}
Here $\eta$ is the ratio in exposure, or ratio of the size of the two regions. For our case of an equally sized on-- and off--source region, $\eta=1$. 

\begin{figure}
\includegraphics[angle=-90,width=3.0in]{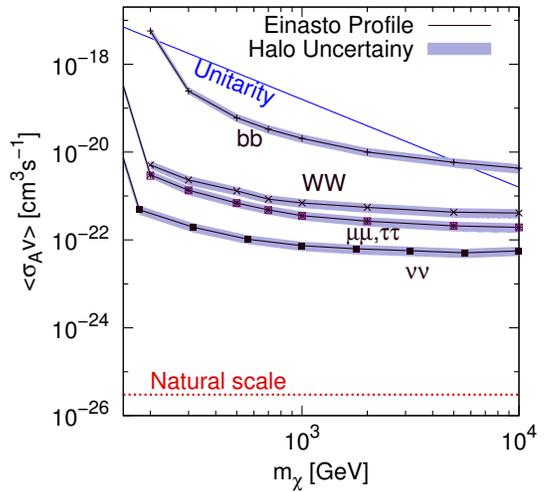}
\caption{(Color online) 90\% C.L. upper limit on the dark matter self annihilation cross section for
  five different annihilation channels. Also shown are the natural scale (red dotted line), for which the WIMP is a thermal relic~\cite{Steigman:1979kw,Jungman:1995df}, and unitarity bound (blue line)~\cite{Griest:1989wd,Hui:2001wy}.
  For the limit curves, the central line is for the Einasto and NFW
  profiles, while the shaded width identifies the extrema results from the Moore and Kravtsov
  profiles. We consider only smooth halo profiles. The limits for $\tau\tau$ and $\mu\mu$ 
 overlay, due to their very similar high energy neutrino spectra.
\label{fig_exclusion_limit}}
\end{figure}

Figure~\ref{fig_exclusion_limit} shows the obtained exclusion limit
compared to the ``natural scale'', for which dark matter candidates are consistent with being a thermal relic~\cite{Steigman:1979kw,Jungman:1995df}. Larger cross sections are possible if, for example, dark matter is produced non-thermally or acquires mass only in the late universe~\cite{Kaplinghat:2000vt}.

\begin{figure}
\includegraphics[angle=-90,width=3.0in]{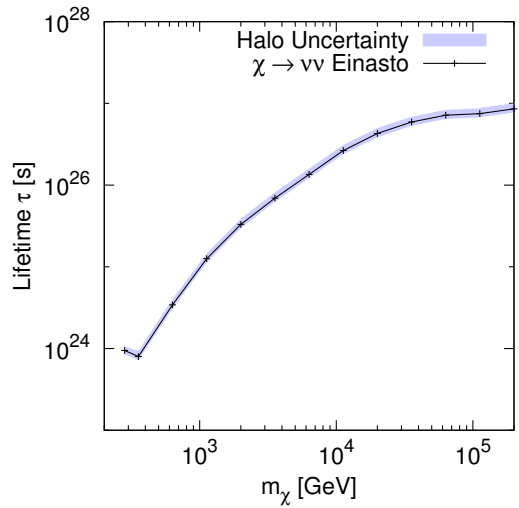}
\caption{
  Lower limit on WIMP lifetime $\tau$ assuming $\chi
  \rightarrow \nu \bar{\nu}$ at $90\%$~C.L..
\label{fig_limit_decay}}
\end{figure}

Applying the same procedure as that above for the annihilation cross section, we compute a 
$90\%$~C.L. lower limit on the WIMP lifetime, $\tau$, as function of
the WIMP mass, as shown in Fig.~\ref{fig_limit_decay}.
We assume a line spectrum, $\chi \rightarrow \nu\nu$ and apply Eq.~\ref{dphidE_decay} 
for the expected neutrino flux. If dark matter is a thermal relic and unstable, the only requirement in order for it to be present today is that it has a lifetime much longer than
the age of the Universe $T_{\rm U} \simeq 4 \times 10^{17}$~s.

\begin{figure}
\includegraphics[width=3.0in]{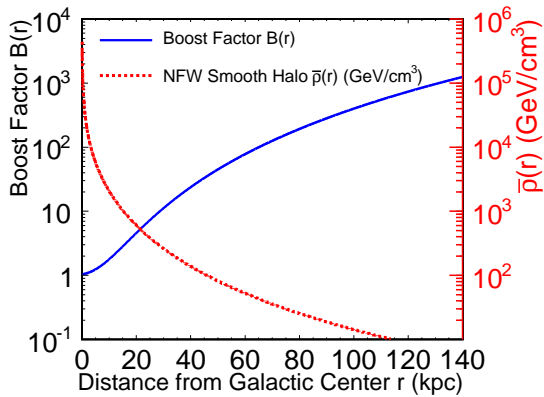}
\caption{
Boost factor as function of the distance from the Galactic Center for the simplest model of~\cite{Kamionkowski:2010mi} and a dark matter density using the NFW halo profile.
\label{fig_sys_boost}}
\end{figure}

Our limit calculation assumes smooth, spherically symmetric halo models.  
However, N-body simulations indicate that dark matter in the halo should have some 
substructure~\cite{Moore:1999nt,Diemand:2006ik}. While this will have negligible effects on the expected neutrino flux from dark matter decay, the presence of substructure will enhance the self-annihilation rate since it is proportional to the square of the dark matter density. 
To quantify the average expected enhancement in the annihilation rate compared to a smooth dark matter distribution,
 one can define a boost factor 
as a function of the distance from the
Galactic Center~\cite{Kistler:2009xf,Kamionkowski:2010mi}:

\begin{equation}
B(r)=\frac{\int \rho^2 dV}{\int(\bar{\rho})^2 dV},
\end{equation}
where we defined $\bar{\rho}$ as the mean density of the smooth halo component.
To determine the impact of a boosted neutrino flux on the expected neutrino
signal in the on-- and off--source regions
we use the signal enhancement resulting from substructure in the halo following the simplest model of reference~\cite{Kamionkowski:2010mi}, as shown in Fig.~\ref{fig_sys_boost}. 

\begin{figure}
\includegraphics[angle=-90,width=2.6in]{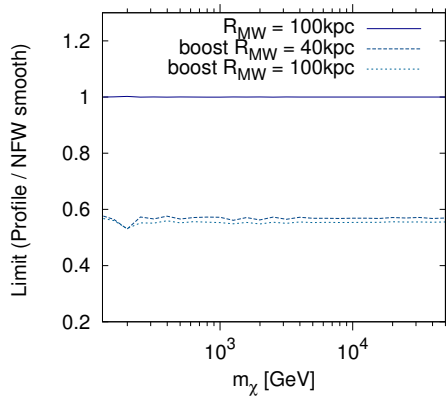}
\caption{
The ratio between the limit obtained with our default and modified halo models are shown. 
The scaling due to a boost factor and the adopted size of the Galactic dark matter halo $R_{\rm MW}$ are given separately.
\label{fig_sys_boost_limits}}
\end{figure}

We investigate the scaling of the limit due to a boost factor and adopted
size of the Galactic dark matter halo, $R_{\rm MW}$, which sets the upper integration limit
in the dark matter density line of sight integral given by Eq.~\ref{Jpsi_integral}. 
The ratio between the limit for the default value (smooth halo, and $R_{\rm MW}=40$~kpc) and the modified halo model is shown in Fig.~\ref{fig_sys_boost_limits}. An increase in the halo size $R_{\rm MW}$ from 40~kpc to 100~kpc has no impact. Boosting the flux due to substructure results in a better limit and therefore assuming no substructure, yields a more conservative result.

Another possible contribution to the neutrino flux from dark matter self-annihilations
originates outside our Galaxy. This extra-galactic flux~\cite{Beacom:2006tt} is
expected to be isotropic and, hence, contributes equally to the on-- and off--source
regions. That is, any such additional flux would equally contribute to the number of events observed
in the on-- and off--source regions and hence make a flux limit based on the
difference more conservative. Note also that the contribution from
the extragalactic component is much smaller
than the flux from within our Galaxy~\cite{Yuksel:2007ac}.

\section{Comparison to phenomenological models}

Lepton signals, such as those observed in the ATIC peak~\cite{ATIC:2008zzr}, the
PAMELA GeV positron excess~\cite{Adriani:2008zr}, and 
electron spectra from H.E.S.S.~\cite{Aharonian:2009ah}, and
Fermi~\cite{Abdo:2009zk} deviate from predictions for the primary electron
and cosmic ray secondary positron spectrum~\cite{Moskalenko:1997gh}. Such an
excess, if interpreted as originating from dark matter self-annihilations, would
be indicative of leptophilic dark matter
candidates~\cite{Meade:2009iu,Cirelli:2008pk}. 
Alternatively, such an excess could also be explained through nearby astrophysical sources such as pulsars~\cite{Yuksel:2008rf}.

\begin{figure*}
\includegraphics[width=3.0in]{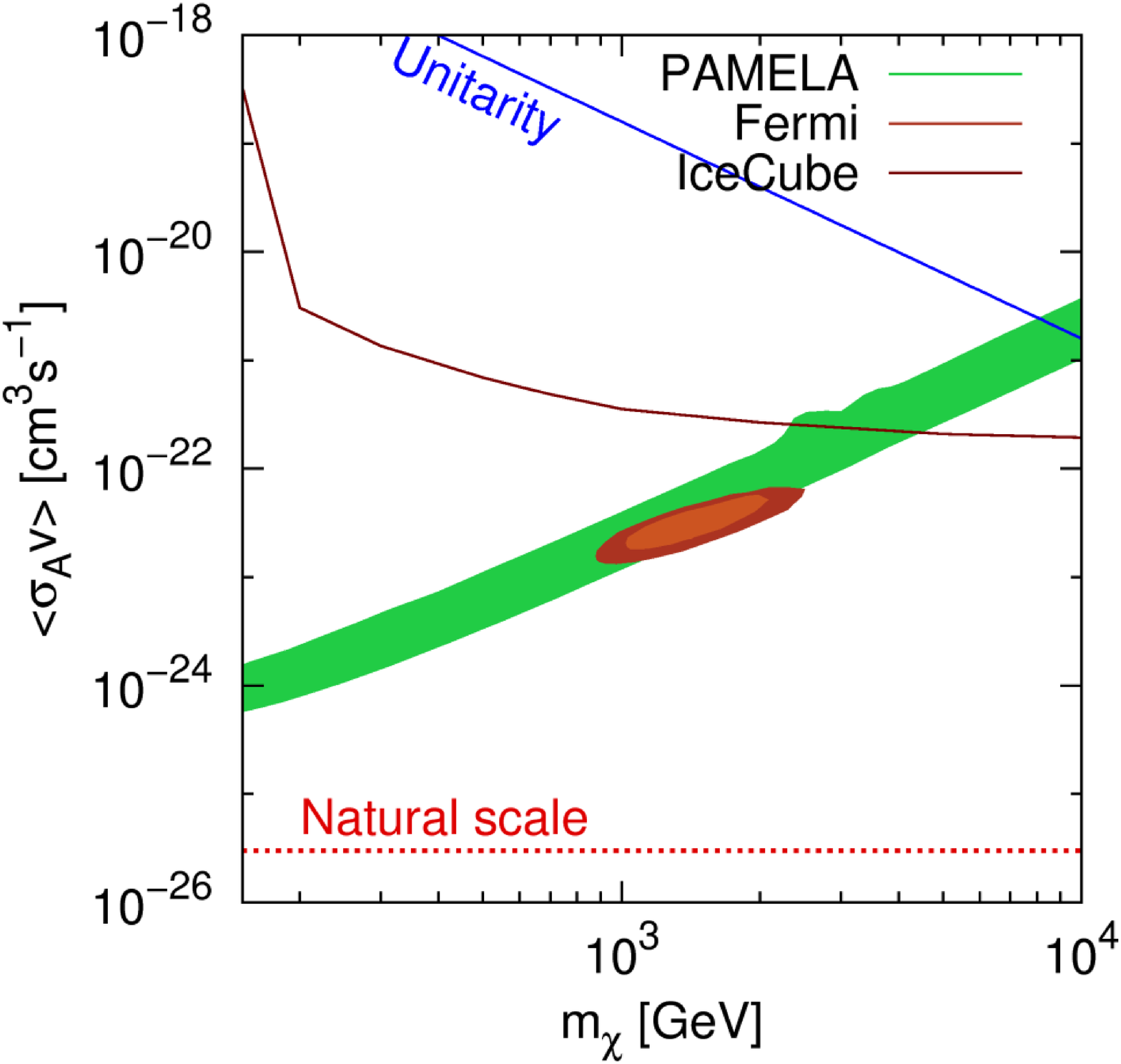}
\includegraphics[width=3.0in]{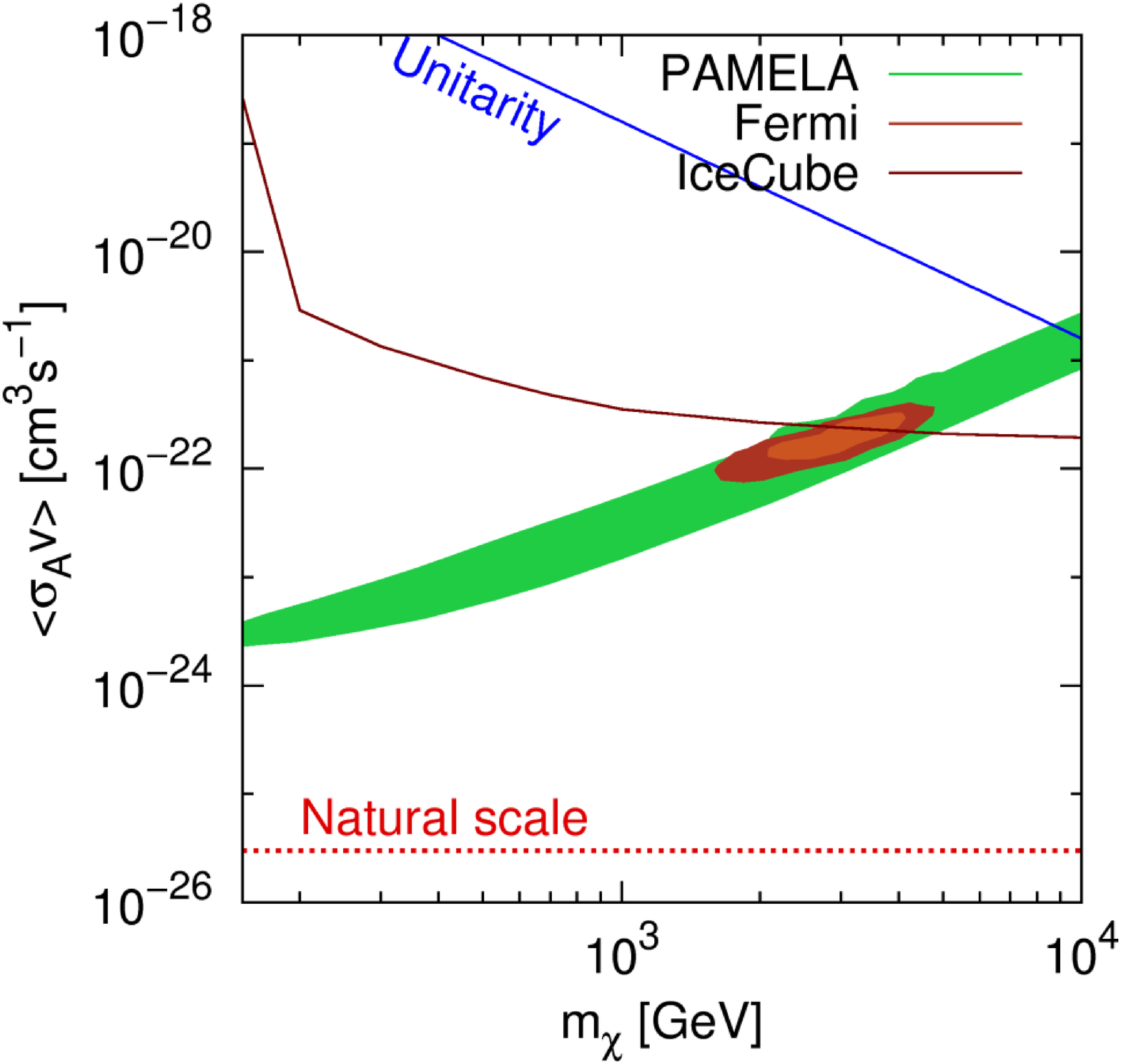}
\caption{(Color online) 90\% C.L. upper limit on the dark matter self annihilation cross section assuming the Einasto profile and annihilation into $\mu\mu$ (left panel) and $\tau\tau$ (right panel). Limits are compared to a preferred phenomenological model to explain the PAMELA excess (green)
together with Fermi electrons (brown). The natural scale (red dotted line), for which the WIMP is a thermal relic, and unitarity bound~\cite{Griest:1989wd,Hui:2001wy} (blue line) are shown.
\label{fig_tautau_fermi}}
\end{figure*}

Since electrons lose significant energy during propagation,
signals must originate within a distance of about one kpc from the Sun.
While electron signals could only probe the local dark matter density, 
the presented large--scale anisotropy search probes a wider range of the 
Milky Way halo.
Figure~\ref{fig_tautau_fermi} compares the
IceCube exclusion limit with phenomenological interpretations of
anomalous electron measurements for two example annihilation 
channels ($\mu\mu$,$\tau\tau$) and our chosen benchmark profile of Einasto. 
Even the small dataset used here allows this analysis to constrain models motivated
by the anomalous lepton signals.

\section{Summary and Outlook}

The IceCube candidate neutrino sample, collected during 2007--2008 in the 22-string
configuration, has been used to search for a neutrino anisotropy as expected from
dark matter self annihilation in the Milky Way halo. Such an anisotropy was
not observed and we have determined limits on the 
dark matter self-annihilation cross section $\langle \sigma_{A} v \rangle$ 
at 90\%~C.L. for WIMPs in the mass range from 200~GeV to 10~TeV.
The IceCube detector sensitivity can be significantly
improved by investigating the Galactic Center  as a potential source. 
Such a search could be performed with the IceCube detector at a later construction stage and rely on selecting neutrinos 
interacting inside the detector volume. It would be able to 
significantly improve the constraints on the dark matter self-annihilation 
cross section given a particular choice of halo model in the case of a non observation.
A large--scale anisotropy study as performed here, however, might provide a more distinct
discovery signal. In the case of the Galactic Center, a dark matter signal 
would be more difficult to distinguish from other astrophysical neutrino sources, such as point sources (source
contamination) or cosmic ray interaction with the interstellar medium. 

\begin{acknowledgments}

We acknowledge the support from the following agencies:
U.S. National Science Foundation-Office of Polar Programs,
U.S. National Science Foundation-Physics Division,
University of Wisconsin Alumni Research Foundation,
the Grid Laboratory Of Wisconsin (GLOW) grid infrastructure at the University of Wisconsin - Madison, the Open Science Grid (OSG) grid infrastructure;
U.S. Department of Energy, and National Energy Research Scientific Computing Center,
the Louisiana Optical Network Initiative (LONI) grid computing resources;
National Science and Engineering Research Council of Canada;
Swedish Research Council,
Swedish Polar Research Secretariat,
Swedish National Infrastructure for Computing (SNIC),
and Knut and Alice Wallenberg Foundation, Sweden;
German Ministry for Education and Research (BMBF),
Deutsche Forschungsgemeinschaft (DFG),
Research Department of Plasmas with Complex Interactions (Bochum), Germany;
Fund for Scientific Research (FNRS-FWO),
FWO Odysseus programme,
Flanders Institute to encourage scientific and technological research in industry (IWT),
Belgian Federal Science Policy Office (Belspo);
University of Oxford, United Kingdom;
Marsden Fund, New Zealand;
Japan Society for Promotion of Science (JSPS);
the Swiss National Science Foundation (SNSF), Switzerland;
A.~Gro{\ss} acknowledges support by the EU Marie Curie OIF Program;
J.~P.~Rodrigues acknowledges support by the Capes Foundation, Ministry of Education of Brazil.

\end{acknowledgments}

\bibliography{HaloPRD_ref_bibtex}

\end{document}